\documentstyle[aps,prl,psfig,twocolumn]{revtex}

\draft

\begin{document}

\title{Hydrodynamical quantum state reconstruction}

\author{Lars M. Johansen \thanks{Email: lars.m.johansen@hibu.no}}
\address{Buskerud College, P.O.Box 251, N-3601 Kongsberg, Norway}
\date{\today}
\maketitle

\begin{abstract}

The density matrix of a nonrelativistic wave-packet in an arbitrary,
one-dimensional and time-dependent potential can be reconstructed by
measuring hydrodynamical moments of the Wigner distribution. An
$n$-th order Taylor polynomial in the off-diagonal variable is
obtained by measuring the probability distribution at $n+1$ discrete
time values.

\end{abstract}
\pacs{PACS number(s): 03.65.Bz, 05.30. d}

This Letter presents a new and general method for reconstructing
the density matrix of a massive particle in an arbitrary,
one-dimensional and time-dependent potential. Such a general method
may seem called for, e.g., in the reconstruction of the quantum state
of particles in anharmonic, time-dependent Paul traps. The method is
based upon measuring the position probability distribution for a
discrete number of time values in a short time interval.
Surprisingly, the method follows almost immediately from known
results.

The diagonal of the density matrix can be retrieved by observing a
single probability distribution. The state reconstruction problem
essentially consists in obtaining the offdiagonal elements.
Decoherence and the approach to the classical regime is characterized
by a vanishing of the offdiagonal elements \cite{Zurek91}. The method
presented here is constructed so that the density matrix is retrieved
for increasing values of the off-diagonal variable by increasing the
number of discrete time values for which the probability density is
observed.

It was shown by Madelung \cite{Madelung26} that quantum mechanics can
be reformulated in a form resembling a hydrodynamical description.
He reformulated the Schr\"odinger equation as two coupled and
nonlinear equations for the ``hydrodynamical" moments of probability
density and probability current density. Whereas these two moments
can describe a pure state \cite{purehydro}, the situation is more
complicated for mixed states.

A somewhat analogous situation is found in classical statistical
mechanics. Hilbert \cite{Hilbert12} demonstrated that the phase space
distribution for a system in local thermal equilibrium can be
expressed as a functional of the density, the current density and the
kinetic energy density. These are the three lowest order velocity
moments of the phase space distribution. Thus, in thermal equilibrium
the phase space distribution is equivalent to it's three lowest order
moments. This situation has sometimes been called the Hilbert paradox
\cite{Uhlenbeck63}. In general, though, an infinite set of velocity
moments is equivalent to the full phase space distribution. These
moments are coupled through an infinite set of differential
equations. In the case of thermal equilibrium this set is truncated,
and a finite and closed set of equations is obtained.

The similarity between quantum mechanics and statistical mechanics
goes beyond the observation made by Madelung, which is restricted to
pure states. Wigner \cite{Wigner32} showed that quantum mechanics can
be reformulated in terms of a quasi phase space distribution. This
distribution, which is the Fourier transform of the density matrix,
shares most of the properties of a classical phase space
distribution. One important exception is that it may take on negative
values. It can be shown \cite{Carruthers83} that the probability
density and the probability current density are the first two
velocity moments of the Wigner distribution. This is in complete
analogy with classical statistical mechanics. An infinite hierarchy
of velocity moments can be derived from the Wigner distribution, and
they are interconnected through an infinite set of coupled equations,
much like in classical statistical mechanics. 

However, quantum mechanics can offer more. It has been shown
\cite{Moyal49,Yvon78,Ploszajczak,Lill} that when the density matrix
in position representation is expanded as a Taylor series in the
offdiagonal variable, the coefficients of this expansion are velocity
moments of the Wigner distribution. This in fact solves the moment
problem both for classical statistical mechanics and for quantum
mechanics. With ``the moment problem" we here mean the problem of
expressing a distribution in terms of it's moments. This is a
classical problem in statistics, and it was first raised in the
context of quantum mechanics by Moyal \cite{Moyal49}. It was further
explored in Ref. \cite{further}. Recently, the density matrix was
expressed in terms of normally ordered moments \cite{normal}. It has
also been shown \cite{calc} that the normally ordered moments can be
calculated from the measured quadrature distribution. Since
convergent expansions may be found \cite{Herzog96}, this gives a
method for reconstructing the state of a radiation field or a
particle in a harmonic oscillator potential.

The possibility of measuring quantum states has attracted a lot of
attention in recent years. In particular, the method of homodyne
tomography \cite{Bertrand87,Vogel89} has contributed to this
interest. It has been used to reconstruct the quantum state of
radiation fields \cite{opthomexp} as well as material particles
propagating in free space \cite{Kurtsiefer97}. In homodyne tomography
the state is retrieved from a parameterized probability distribution.
Ideally, a continuous range of parameter values should be used. In
the case of optical homodyne tomography, this parameter is the value
of a reference phase \cite{Vogel89}, whereas for material particles
it might be a time value \cite{Raymer94b,Leonhardt96}. In another
recently developed reconstruction scheme, a method for the direct
probing of the Wigner distribution has been found \cite{direct}. For
a specific parameter value (in this case, the amplitude and phase of
a probe field) a certain region of phase space is retrieved. This
method has been used to reconstruct the first negative Wigner
distribution \cite{Leibfried96}. Numerous other reconstruction
methods have also been found \cite{survey}.

Recently, it has been shown that the density matrix of a wave-packet
in an arbitrary one-dimensional potential can be reconstructed by
observing the time-evolution of the position probability density
\cite{Leonhardt96,Opatrny97}. In these methods, the eigenstates of
the Schr\"odinger equation are first found for the potential in
question. The position probability density should ideally be observed
over an infinite time interval, although methods have been considered
for obtaining a finite observation time \cite{Opatrny97,Leonhardt97}.

The density matrix is the Fourier-transform of the Wigner
distribution \cite{Wigner32}
\begin{equation}
	\langle x + y | \, \hat{\rho} \, | x - y \rangle =
	\int_{-\infty}^{\infty} dp \, e^{2 i p y/\hbar} \, W(x,p,t).
	\label{eq:onedimFourier}
\end{equation}
It follows that \cite{Moyal49}
\begin{equation}
	\left [ {\partial^{(n)} \langle x + y | \, \hat{\rho} \, | x - y
	\rangle \over \partial y^n} \right ]_{y=0} = \left ( {2 i \over
	\hbar} \right )^n \, f_n(x,t),
	\label{eq:coeff}
\end{equation}
where $f_n$ are ``hydrodynamical" moments of the Wigner distribution
\cite{Moyal49}
\begin{equation}
	f_n (x,t) = \int_{-\infty}^{\infty} dp \, p^n \, W(x,p,t).
	\label{eq:moments}
\end{equation}
Since the Wigner distribution is real, these moments are also real.
$f_0$ trivially is the probability distribution in position
representation, whereas $f_1/m$ is the probability current density
\cite{Carruthers83}. Generally it is not possible to give these
moments a classical hydrodynamical interpretation. Thus, e.g., the
moment $f_2$ may take on negative values for certain negative Wigner
distributions \cite{Johansen97b}.

It follows that the unique Taylor expansion of the density matrix in
the off-diagonal variable $y$ is \cite{Yvon78,Ploszajczak,Lill}
\begin{equation}
	\langle x + y | \, \hat{\rho} \, | x - y \rangle =
	\sum_{n=0}^{\infty} {f_n(x,t) \over n!} \, \left ({2 i y \over
	\hbar} \right )^n.
	\label{eq:series}
\end{equation}
We may divide this expansion into a real and an imaginary part by
\begin{eqnarray}
	\langle x + y | \, \hat{\rho} \, | x - y \rangle =
	\sum_{n=0}^{\infty} (-1)^n {f_{2n}(x,t) \over (2n)!} \, \left
	({2 y \over \hbar} \right )^{2n} \nonumber \\ + \, i \:
	\sum_{n=0}^{\infty} (-1)^n {f_{2n+1}(x,t) \over (2n+1)!} \,
	\left ({2 y \over \hbar} \right )^{2n+1}.
	\label{eq:complexform}
\end{eqnarray}
We see that the real part contains only moments $f_n$ of even order
$n$, whereas the imaginary part contains only moments of odd order.

Clearly, if we are able to measure the moments $f_n$, we have a state
reconstruction scheme. To this end, we recall the equation of
motion of the Wigner distribution for a particle with mass $\mu$ in a
one-dimensional, time-dependent potential $V(x,t)$,
\cite{Wigner32,Lill}
\begin{eqnarray}
	{\partial \over \partial t} W(x,p,t) &=& \left \{ - {p \over
	\mu} {\partial \over \partial x} + \sum_{k=0}^{\infty} \left (
	{\hbar \over 2i} \right )^{2k} {1 \over (2k+1)!} \right .
	\nonumber \\ &\times& \left . {\partial^{2k+1} V(x,t) \over
	\partial x^{2k+1}} {\partial^{2k+1} \over \partial p^{2k+1}}
	\right \} W(x,p,t).
	\label{eq:wigeqmot}
\end{eqnarray}
We multiply both sides with $p^n$ and integrate over all momentum
space. In this way, we obtain the infinite set of coupled equations
\cite{Ploszajczak,Lill}
\begin{eqnarray}
	{\partial f_n \over \partial t} &=& -{1 \over \mu} {\partial
	f_{n+1} \over \partial x} \nonumber \\ &-& \sum_{k=0}^{[(n-1)/2]}
	\left ( {\hbar \over 2i} \right )^{2k} \left ( \begin{array}{c} n
	\\ 2k+1 \end{array} \right ) {\partial^{2k+1} V \over \partial
	x^{2k+1} } \: f_{n-2k-1}.
	\label{eq:conserv}
\end{eqnarray}
For $n=0$ we retrieve the well known conservation equation for 
probability. By integrating this conservation equation, the
probability current density can be expressed in terms of the
time-derivative of the cumulative position probability \cite{Royer89}
(for one-dimensional systems, that is). The idea of the present
reconstruction method is simply to generalize this procedure to
arbitrary moments. This will yield an iterative scheme. We therefore
integrate Eq. (\ref{eq:conserv}) over the position variable and
obtain \cite{classical}
\begin{eqnarray}
	f_{n+1}(x,t) &=& - \mu {\partial \over \partial t}
	\int_{-\infty}^x dx' \, f_n(x',t) \nonumber \\ &-& \, \mu
	\sum_{k=0}^{[(n-1)/2]} \left ( {\hbar \over 2i} \right
	)^{2k} \left ( \begin{array}{c} n \\ 2k+1 \end{array} \right )
	\nonumber \\ &\times& \int_{-\infty}^x dx' \, {\partial^{2k+1}
	V(x',t) \over \partial x'^{2k+1}} f_{n-2k-1}(x',t).
	\label{eq:recursive}
\end{eqnarray}
The moment $f_{n+1}$ is expressed in terms of lower order moments
only. Therefore an arbitrary moment can be recursively calculated
from the zeroth order moment, the probability density. This recursion
relation gives an algorithm for reconstructing the density matrix.
The algorithm can be used directly on the experimental data. In 
order to find $f_n$, we must know the time derivative of $f_{n-1}$.
Therefore, $f_{n-1}$ must be observed for at least two different time
values. This again requires that $f_{n-2}$ is known for three time
values. Recursively, it follows that $f_n$ can be found by measuring
$f_0$ for at least $n+1$ different time values.

\begin{figure}
		\begin{footnotesize}
        \centerline{\psfig{figure=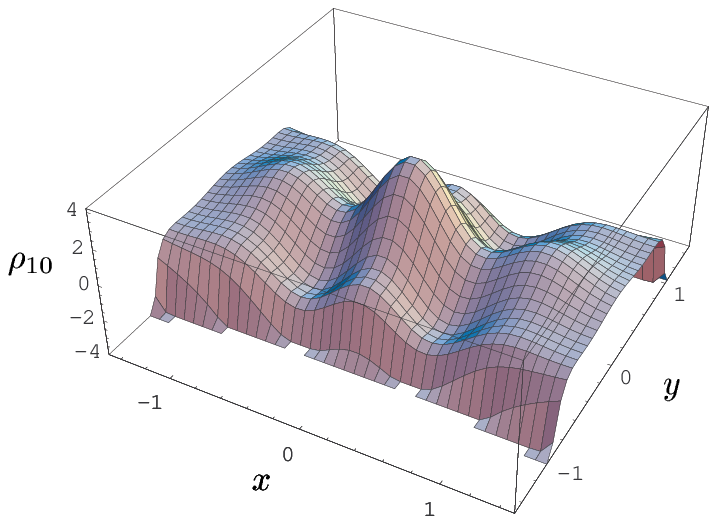,width=7cm}}
		\bigskip
		\centerline{a)}
		\bigskip
        \centerline{\psfig{figure=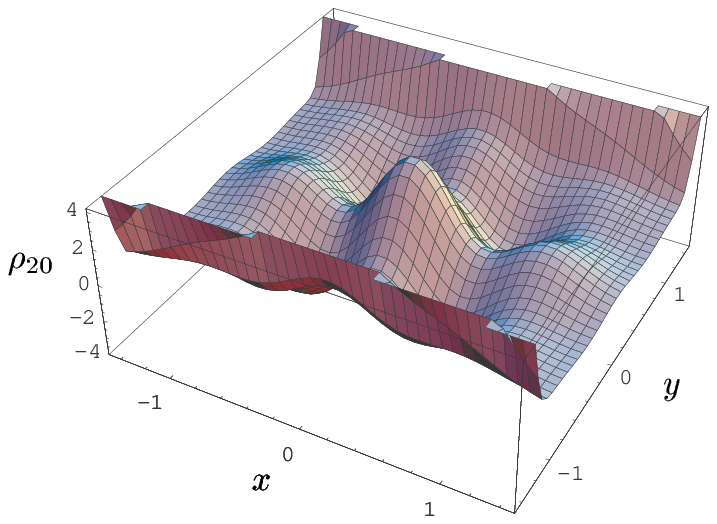,width=7cm}}
		\bigskip
		\centerline{b)}
		\bigskip
        \centerline{\psfig{figure=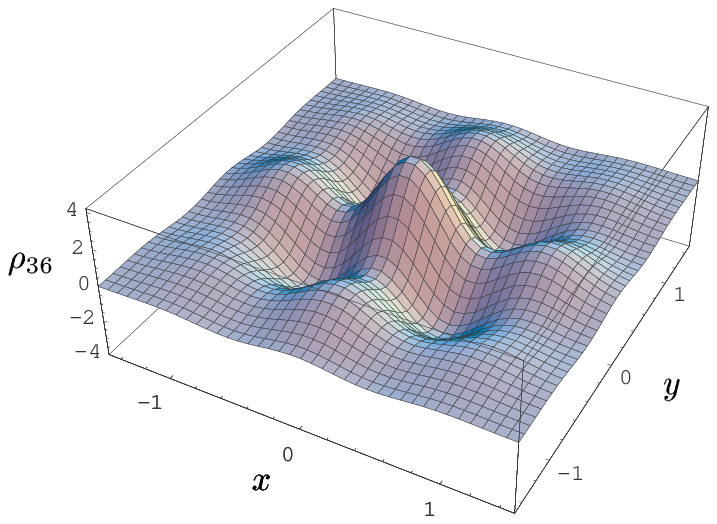,width=7cm}}
		\bigskip
		\centerline{c)}
		\bigskip
        \caption{The real part of the Taylor polynomial $\rho_N$ of
        the density matrix for a) $N=10$ b) $N=20$ and c) $N=36$. It
        is seen that the density matrix is retrieved for increasing
        values of the offdiagonal variable as the number of discrete
        time values are increased. $\rho_{36}$ is almost
        indistinguishable from the exact density matrix within the
        chosen plotting region.}
        \label{fig:density}
		\end{footnotesize}
\end{figure}

To illustrate the convergence of the Taylor series (\ref{eq:series}),
consider the unnormalized superposition state
\begin{equation}
	\psi(x) = e^{- [x/(2 \sigma)]^2 + i k_0 x } + e^{- [x/(2
	\sigma)]^2 - i k_0 x }.
\end{equation}
The corresponding density matrix is
\begin{eqnarray}
	\langle x+y | \, \hat{\rho} \, | x-y \rangle &=& 2 \exp \left [ -
	{x^2 + y^2 \over 2\sigma^2} \right ] \nonumber \\ &\times& \left
	[ \, \cos (2 k_0 x) + \cos (2 k_0 y) \, \right ].
\end{eqnarray}
Note that this density matrix is real. This means, according to Eq.
(\ref{eq:complexform}), that $f_n$ vanishes for all odd $n$. In Fig.
\ref{fig:density} the Taylor polynomial 
\begin{equation}
	\rho_N(x,y,t) = \sum_{n=0}^{N} {f_n(x,t) \over n!} \, \left ( {
	2 i y \over \hbar} \right )^n 
	\label{eq:polynomial}
\end{equation}
has been plotted for different orders $N$ for the parameter choice
$\sigma=1/\sqrt{2}$ and $k_0=2 \sqrt{2}$. The highest order
polynomial is $\rho_{36}$ (Fig. \ref{fig:density} c), which involves
moments up to $f_{36}$. It would be obtained by measuring the
position probability distribution $f_0$ for 37 different time values
using perfect detectors. It differs negligibly from the exact density
matrix within the chosen plotting region.

As we can see from Fig. \ref{fig:density}, one in effect probes the
density matrix further away from the diagonal by increasing the
number of discrete time values for which the probability distribution
is observed. If, after retrieving the Taylor polynomial
(\ref{eq:polynomial}) to a certain order, one finds that the density
matrix goes to zero, one may use an additional number of measurements
at other time values to check the consistency of the data.

The classical limit is often associated with taking $\hbar
\rightarrow 0$. In this case the equation of motion
(\ref{eq:wigeqmot}) reduces to a classical Liouville equation. But
for the purpose of reconstructing  the Taylor polynomial
(\ref{eq:polynomial}), it is vital that $\hbar$ should be considered
finite. Otherwise, assuming finite hydrodynamical moments $f_n$,
every term in the polynomial of order higher than zero diverges. The
specific numerical value of $\hbar$ is of no importance in this
respect, since a change of $\hbar$ only implies a rescaling of the
off-diagonal variable.

The principles outlined in this Letter can also be employed to other
expansions of the density matrix. Starting with a density matrix in
the momentum representation, we might have expanded it in terms of
the moments $\int_{-\infty}^{\infty} dx \, x^n \, W(x,p,t)$. However,
the corresponding set of recursion relations is generally more
complicated in this case. Other expansions of the density matrix,
which converge more rapidly for nearly classical states \cite{Lill},
might also be used for state reconstruction using similar techniques.
It is also easily adapted in other areas such as the reconstruction
of quantum optical states.

The expansion (\ref{eq:series}) of the density matrix in terms of
quasi-hydrodynamical moments can be generalized to systems with a
higher number of dimensions. However, it may not be straightforward
to generalize the recursion algorithm (\ref{eq:recursive}). This is
basically due to the fact that the current density is not uniquely
determined from the time derivative of the probability density for
systems with two or more dimensions.

In conclusion, a method was found for reconstructing the density
matrix of a particle in an arbitrary, time-dependent potential. The
method was based upon a Taylor expansion of the density matrix in the
off-diagonal variable. The coefficients $f_n$ in this expansion are
hydrodynamical moments of the Wigner distribution. A recursive
algorithm was found for calculating an arbitrary moment $f_n$ from
the zeroth order moment, the probability distribution. In general, an
$n$-th order Taylor polynomial of the density matrix can be found by
observing the probability distribution at $n+1$ discrete time values.


\begin{thebibliography}{10}
\bibitem{classical} It can be noted that the recursion relations
(\ref{eq:recursive}) coincide with the corresponding classical
relations obtained from the classical Liouville distribution provided
that the potential is a polynomial of maximally second order in $x$.
Moreover, the recursion relations for $n=0,1,2$ (giving $f_1$,
$f_2$ and $f_3$) coincide with the classical ones regardless of
potential.
\end{thebibliography}
\end{document}